\documentclass[prl,twocolumn,superscriptaddress,floatfix,longbibliography]{revtex4}
\usepackage{enumerate}
\usepackage{graphicx}
\usepackage{amsmath}
\usepackage{amsthm}   
\usepackage{mathtools}
\usepackage{xcolor}
\usepackage{amsfonts}
\usepackage{amssymb}
\usepackage{units}
\usepackage[normalem]{ulem}
\newcommand{\ket}[1]{| #1 \rangle}
\newcommand{\bra}[1]{\langle #1 |}
\newcommand{\Tr}{\text{Tr}}

    \newtheorem*{theorem}{Theorem}
    
\begin{document}
    
\title{Energy Ordering from Nonlinear Quantum Dissipation}

\author{Alireza Ataei\footnote{Electronic address: alireza.ataei@math.uu.se}}
\affiliation{Department of Mathematics, Uppsala University, Uppsala, Sweden}

\author{Olle Eriksson}
\affiliation{Department of Physics and Astronomy, Uppsala University, Box 516, 
SE-751 20 Uppsala, Sweden}
\affiliation{Wallenberg Initiative Materials Science,
WISE, Uppsala University, Box 516, 
SE-751 20 Uppsala, Sweden}

\author{Vahid Azimi-Mousolou\footnote{Electronic address: vahid.azimi-mousolou@physics.uu.se}}
\affiliation{Department of Physics and Astronomy, Uppsala University, Box 516, 
SE-751 20 Uppsala, Sweden}

    \date{\today}

    \begin{abstract}

Energy-ordered occupation is deeply embedded in quantum physics, from the Aufbau principle governing the filling of electronic states in atoms and molecules to the emergence of low-energy configurations in quantum many-body systems. 
However, the dynamical mechanism by which a generic quantum state develops such an energy hierarchy remains a fundamental question.
Here we show that such an energy hierarchy can emerge dynamically from nonlinear quantum dissipation. 
Rather than being imposed as a principle or generated through coupling to a thermal reservoir, an Aufbau-like ordering of energy levels emerges intrinsically under quantum Landau-Lifshitz-Gilbert dynamics from a generic initial mixed state.
This convergence is a nontrivial consequence of Lyapunov monotonicity and instability of disordered population configurations. 
The resulting dynamics establish an intrinsic nonlinear mechanism for organizing density-matrix populations and provides a route toward selective preparation of low-energy subspaces. Numerical simulations confirm analytical predictions and illustrate convergence toward low-energy sectors.

    \end{abstract}
    
    \maketitle
\textit{Introduction}--A fundamental question across quantum many-body physics, quantum chemistry, condensed matter, and quantum information science is how, during quantum dynamics, quantum populations reorganize themselves with respect to the underlying energy spectrum. Energy provides a natural hierarchy of quantum states, with low-energy configurations playing a central role in determining the properties of molecules, materials, and interacting many-body systems. In quantum chemistry, this hierarchy is reflected in the Aufbau principle, according to which lower-energy orbitals are preferentially occupied before higher-energy ones \cite{Slater1930}. Together with the Pauli exclusion principle and Hund's rules, the Aufbau principle provides a familiar organizing picture for electronic configurations and indeed the form of the Periodic Table. However, the dynamical origin of such ordering remains an open question. A natural route to such dynamical organization is dissipation, which is studied here in detail.

Dissipation is a powerful mechanisms for describing relaxation and self-organization in physical systems. In magnetic materials, the Landau--Lifshitz--Gilbert (LLG) equation describes the damped dynamics of magnetization and provides a mechanism by which a magnetic moment relaxes toward energetically favorable configurations \cite{Landau1935,Gilbert1955,Stiles2006,Ralph2008,Antropov1995}. More generally, dissipative dynamics plays a central role in relaxation, optimization, and state selection across physics, from classical magnetic dynamics to open quantum systems and dissipative state engineering \cite{Breuer2007,Diehl2008,Verstraete2009}. Dissipation therefore offers a natural physical mechanism through which an initially disordered configuration can be reorganized according to an underlying energy landscape.

Recently, a quantum analogue of the LLG equation (QLLG), was introduced as a nonlinear dissipative evolution for quantum spin states \cite{Liu2024}. 
In contrast to conventional open-system descriptions \cite{Lindblad1976,Breuer2007,Gorini1976}, QLLG generates relaxation through an intrinsic nonlinear dynamics while preserving the spectrum of the density matrix. 
In a recent work \cite{Ataei2026}, we demonstrated that QLLG drives generic pure states toward the lowest-energy eigenstate contained in their initial support, providing a physically realizable mechanism for ground-state preparation and quantum optimization. 

 Ground-state vector preparation in \cite{Ataei2026} concerns pure states, whereas realistic quantum systems are generally described by mixed states \cite{Nielsen2010,Breuer2007}. The central object is then no longer a state vector but a density matrix with a nontrivial eigenvalue spectrum. The key question is then what principle governs the dissipative evolution of such quantum populations? More specifically, if QLLG decreases the energy while preserving the spectrum of the density matrix, how are the occupation probabilities redistributed among the energy eigenstates?

To the best of our knowledge, neither classical LLG nor existing analyses of QLLG provide an answer to this question. Classical LLG describes the relaxation of a magnetization vector and therefore provides no framework for characterizing density-matrix spectra, population redistribution, or the ordering of mixed quantum states. On the other hand, the pure-state QLLG evolution addresses only a single occupied quantum state dynamics, leaving mixed states evolution under nonlinear quantum dissipation essentially unexplored.

In this work, we prove that generic finite-rank mixed states evolve toward a unique asymptotic configuration that is diagonal in the energy eigenbasis. More explicitly, we find that the resulting ordering is analogous to the Aufbau principle of atomic and molecular physics, where lower-energy eigenstates are preferentially occupied before higher-energy ones \cite{Slater1930}. Here, however, the ordering emerges dynamically from nonlinear quantum dissipation rather than being imposed as a rule based on minimizing the total energy. From a mathematical perspective, the resulting ordering is closely related to rearrangement and majorization principles that connect ordered spectra to extremal energy configurations \cite{Hardy1929,Marshall2010,Bhatia1997}. Using a Lyapunov characterization of the QLLG equation, together with a stability analysis of stationary density matrices, we show that any diagonal configuration violating the population-energy hierarchy is unstable. Consequently, the Aufbau-ordered state emerges as the unique stable asymptotic arrangement compatible with the conserved spectrum of the density matrix.

Numerical simulations of Heisenberg spin chains confirm analytical results and reveal that the efficiency of the ordering process is controlled by the spectral structure of the Hamiltonian. In particular, the relevant energy gaps govern both the convergence rate and the asymptotic accuracy, linking dissipative performance directly to spectral properties. These results extend QLLG beyond ground-state preparation and establish a general theory of dissipative population sorting in mixed-state quantum systems. This suggests potential applications in low-energy subspace preparation, quantum optimization, state engineering, and dissipative quantum information processing. 

\textit{Dissipative quantum state selection}--The recently proposed quantum Landau-Lifshitz-Gilbert (QLLG) equation \cite{Liu2024}, which  describes nonlinear dissipative dynamics, is not only a quantum analog of classical counterpart typically used to describe the magnetization dynamic of magnetic materials \cite{Landau1935,Gilbert1955} but also offers an approach for quantum optimization \cite{Ataei2026}. The QLLG equation is given by  
\begin{equation}
    \label{eq:QLLG equation}
    \dot \rho = \frac{i}{\hbar}[\rho,H] + i \kappa [\rho, \dot \rho],
    \end{equation}
    where $H$ is the  Hamiltonian of the quantum system and $\kappa>0$ is the damping parameter. 
    The density matrix $\rho(t)$ evolves according to Eq.~\eqref{eq:QLLG equation} from an initial state $\rho_0$.
    
    While the dynamical behavior of QLLG is relatively well understood in the pure-state case \cite{Ataei2026}, the dynamics of mixed states under the QLLG equation remain unexplored. In this work, we analyze the generic dynamic of QLLG for an arbitrary rank initial mixed state, and we prove the following result. 

\begin{theorem}
\label{thm:main}
Under QLLG dynamics, any finite-rank mixed quantum state evolves toward a stationary state that is diagonal in the energy eigenbasis. Moreover, in the asymptotic state, lower-energy eigenstates having nonzero overlap with the initial state acquire larger probabilities. Thus, the dynamics intrinsically induces a stable redistribution of quantum populations analogous to an Aufbau-like filling of the quantum states.
\end{theorem}

This theorem states that, for a random mixed initial state of rank $l$,
\begin{equation}
\rho_0 = \sum_{i=1}^l p_i \ket{\psi_i}\bra{\psi_i},
\label{Ninitialstate}
\end{equation}
where
\begin{equation}
p_1 \ge p_2 \ge \cdots \ge p_l > 0,
\qquad
\sum_{i=1}^l p_i = 1,
\label{eq:normalization of initial state}
\end{equation}
and $\{\ket{\psi_i}\}_{i=1}^l$ is a set of linearly independent states in a $d$-dimensional Hilbert space $\mathcal{H}$, with $l \le d$, the QLLG dynamics asymptotically drives the system toward
\begin{equation}
\lim_{t \rightarrow \infty} \rho(t)
= \rho_{\textup{Auf}}
:= \sum_{i=1}^l p_i \ket{E_i}\bra{E_i}.
\label{Aufbaustate}
\end{equation}
Here, the energy eigenstates $\ket{E_i}$ of the Hamiltonian $H$, which acts on $\mathcal{H}$, 
are assumed to be in nondecreasing order with possible degeneracies,

\begin{equation}
H \ket{E_i} = E_i \ket{E_i},
\qquad
E_1 \le E_2 \le \cdots \le E_d.
\label{GHamiltonian}
\end{equation}
Note that the resulting Aufbau state $\rho_{\textup{Auf}}$ in Eq.~\eqref{Aufbaustate} minimizes the average energy $\Tr(\rho H)$ over the corresponding isospectral manifold; see the Supplementary Materials \cite{Supplementary Materials}.

We prove the theorem in two steps. 
First, we show that the QLLG dynamics dissipates the system toward a mixed state that commutes with the Hamiltonian. In particular, this proves that the asymptotic limit state is diagonal in the energy eigenbasis. 
    Second, we prove a hierarchical ordering of the energy-state probabilities, showing that the equilibrium state reached under QLLG dynamics increasingly concentrates on the lowest-energy eigenstates, in the sense that lower-energy eigenstates are assigned higher occupation probabilities, while the rank is preserved.
      \begin{figure*}[t]
    \centering
\includegraphics[width=0.9\textwidth]{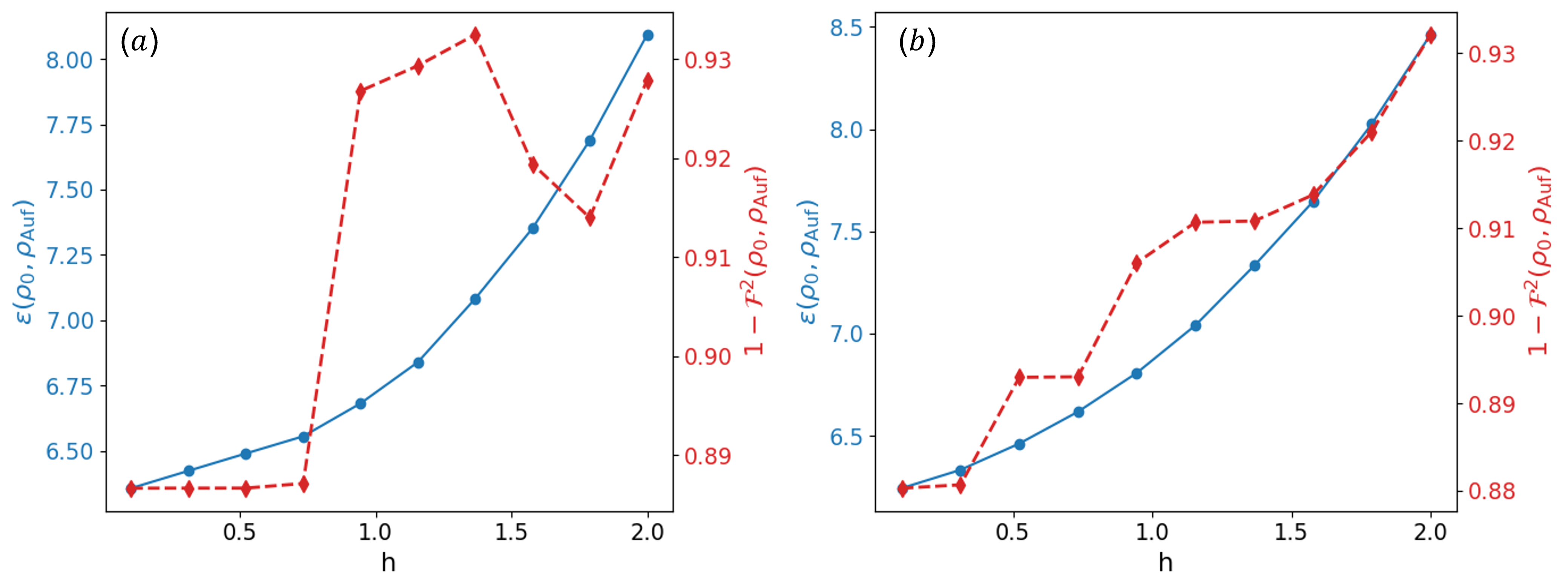}
    \caption{Infidelity and residual energy error between the initial random state and the Aufbau state as a function of magnetic field strength $h$. The residual energy error and $h$ are expressed in units of $J$ and $J/\mu$, respectively, while the infidelity is dimensionless.}
\label{fig:initialerrro}
    \end{figure*} 
    
    \textit{Step 1. Lyapunov Structure and Convergence }-- Considering the energy eigenvalue equation in Eq.~\eqref{GHamiltonian}, define the following energy functional
    \begin{equation}
    V(\rho) = \mathrm{Tr}(\rho H) - E_1 .
    \end{equation}
    By using the QLLG equation, one verifies
    
    \begin{equation}
    \dot V \le 0 ,
    \end{equation}
    for $\kappa < \frac{1}{4}$
    with equality if and only if \cite{Ataei2026}
    \begin{equation}
    [\rho,H]=0.
    \end{equation}
    Since the QLLG flow preserves the spectrum of $\rho$ \cite{Liu2024}, $\rho(t)$ converges to a stationary state $\rho(\infty)$ that commutes with $H$.
    
    \textit{Step 2. Hierarchical Occupation of Energy States}--
    Here, we complete the proof by showing that the equilibrium point $\rho(\infty)$ is unstable under QLLG dynamics if 
    \begin{align}
    \label{eq:hierachydensity}
         E_j > E_i\ \  \text{and}\ \  \rho_{jj}(\infty) >   \rho_{ii}(\infty),
    \end{align}
    for at least a pair of indices $i<j$. To prove this, consider the anti-Hermitian generator
    \begin{equation}
    A_{ij} := \ket{E_i}\bra{E_j} - \ket{E_{j}}\bra{E_{i}}
    \end{equation}
    and the unitary transformation
    \begin{equation}
    U_\epsilon := e^{\epsilon A_{ij}}, \quad
    \rho_{\epsilon} := U_\epsilon \rho(\infty) U_\epsilon^\dagger,
    \end{equation}
    for a real parameter $\epsilon$. By expanding $\rho_{\epsilon}$ around $\rho(\infty)$ with small $\epsilon$, we arrive at
    \begin{equation}
    \rho_{\epsilon} = \rho(\infty)
    + \epsilon [A_{ij},\rho(\infty)]
    + \frac{\epsilon^2}{2}[A_{ij},[A_{ij},\rho(\infty)]]
    + O(\epsilon^3),
    \end{equation}
   where it follows from {\it Step 1} that 
 
    \begin{equation}
    \begin{aligned}
    [A_{ij},\rho(\infty)]
    =&
    (\rho_{ii}(\infty)-\rho_{jj}(\infty))\times
    \\&\left(\ket{E_{i}}\bra{E_{j}} + \ket{E_{j}}\bra{E_{i}}\right)
   \\
    [A_{ij},[A_{ij},\rho(\infty)]]
   =& 
    2(\rho_{ii}(\infty)-\rho_{jj}(\infty))
    \times \\&\left(\ket{E_{i}}\bra{E_{i}} - \ket{E_{j}}\bra{E_{j}}\right)
    \end{aligned}
    \end{equation}
    Hence, the energy difference becomes
    \begin{equation}
    \begin{aligned}
    \delta E&=\Tr(\rho_{\epsilon} H)- \Tr(\rho(\infty) H)
    \\&=
    -\epsilon^2
    (\rho_{jj}(\infty)-\rho_{ii}(\infty))
    (E_{j}-E_{i})
    +
    O(\epsilon^3).
    \end{aligned}
    \end{equation}
    Since both factors of this expression are positive, $\delta E<0$. In conclusion, the QLLG dynamics starting from $\rho_{\epsilon}$ do not converge to $\rho(\infty)$ for small $\epsilon$. Hence, although $\rho(\infty)$ is a stationary state, it is an unstable equilibrium under the condition in Eq.~\eqref{eq:hierachydensity}. This completes the proof, establishing that the asymptotic state is diagonal in the energy eigenbasis and takes the form of the Aufbau state,
\begin{equation}
\label{eq:aufbaustationary}
\rho(\infty) = \rho_{\textup{Auf}}
= \sum_i p_i \ket{E_i}\bra{E_i},
\end{equation}
where $\{p_i\}$ and $\{\ket{E_i}\}$ are the eigenvalues of $\rho_0$ and the energy eigenstates of the Hamiltonian, respectively, and are arranged according to Eqs.~\eqref{eq:normalization of initial state} and \eqref{GHamiltonian}.

    \begin{figure*}[t]
    \centering
    \includegraphics[width=0.9\textwidth]{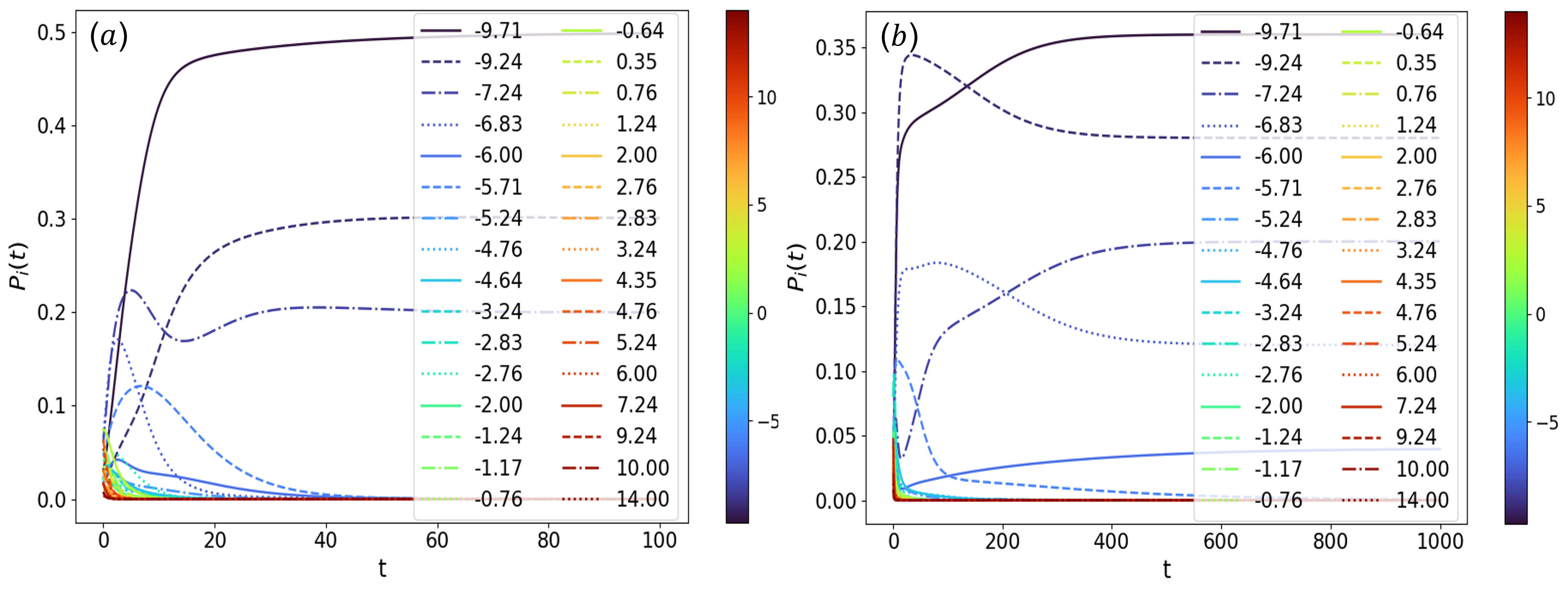}
    \includegraphics[width=0.9\textwidth]{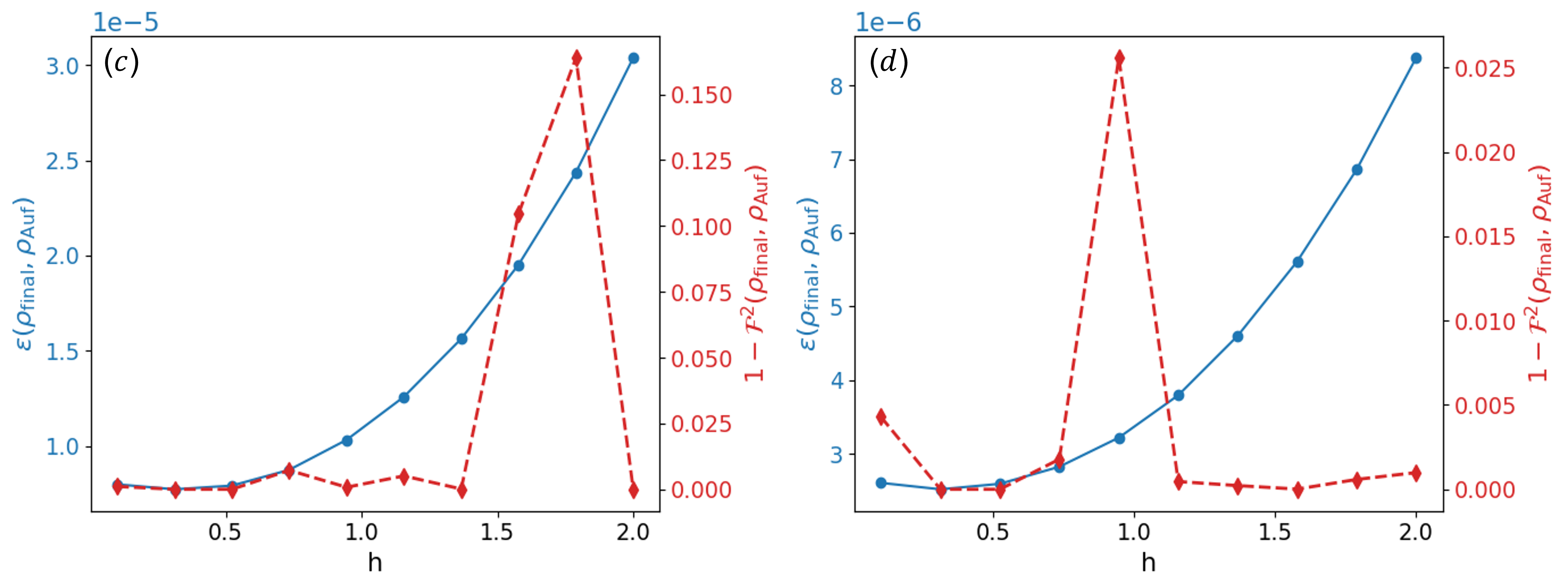}
    \caption{Numerical simulations of QLLG dynamics converging to the Aufbau
state for a five-spin Heisenberg chain. Panels (a) and (b) show the probability
energy-level populations $P_i(t)$ for rank-$3$ and rank-$5$ initial mixed
states, respectively, at $h=2$. In each panel, the line styles and colors
identify the corresponding energy eigenvalues, as indicated in the legend.
Panels (c) and (d) show the residual energy error, normalized by the
exchange interaction $J$, and the infidelity between the final QLLG state
and the Aufbau state as functions of the transverse field $h$. The
magnetic field $h$ and time $t$ are expressed in units of $J/\mu$ and $\hbar/J$,
respectively.}
    \label{fig:population and final error}
    \end{figure*}

\textit{Numerical Simulations}--We illustrate the analytical results of the theorem using a finite Heisenberg spin-$\frac{1}{2}$ chain subject to a transverse magnetic field. The Hamiltonian is given by
\begin{equation}
\label{eq:Heisenbergspinchain} 
H
=
\sum_{i=1}^{N-1}
J\boldsymbol{\sigma}_i \cdot \boldsymbol{\sigma}_{i+1}
- \mu
h \sum_{i=1}^{N} \sigma_i^x.
\end{equation}
Here $J$ denotes the exchange interaction strength, $h$ is the external magnetic field, and $\mu$ is the magnetic moment. In all simulations, we consider an antiferromagnetic regime and set $N=5, J=1$, $\mu=1$, $\hbar=1$, and $\kappa=0.2$. These values are selected as model parameters to demonstrate the theorem above. However, the method and theoretical framework are applicable to general physical systems with arbitrary parameter choices. In realistic magnetic systems, the Heisenberg exchange interaction would typically be expected to dominate over the applied field strength.

The initial conditions are chosen as generic finite-rank mixed quantum states of the form given in Eqs.~\eqref{Ninitialstate} and \eqref{eq:normalization of initial state}. We consider two cases with ranks $l=3$ and $l=5$, where the states $\{\ket{\psi_i}\}_{i=1}^l$ are orthonormal and generated by applying a Haar-random unitary transformation. For the rank-$3$ case, we choose the probability weights
\begin{equation}
\label{eq:rank3overlaps}
p_1=0.5,\qquad p_2=0.3,\qquad p_3=0.2,
\end{equation}
 and, for the rank-$5$ case, we choose
\begin{equation}
\label{eq:rank5overlaps}
\begin{aligned}
&p_1=0.04,\qquad p_2=0.12,\qquad p_3=0.20,\\
&p_4=0.28, \qquad p_5=0.36.
\end{aligned}
\end{equation}
 This construction generates rank $3$ and $5$ density matrices on $2^5$-dimensional Hilbert space.

 Figure \ref{fig:initialerrro} shows the initial residual energy error 
\begin{equation}
\varepsilon(\rho_0,\rho_{\textup{Auf}} )= \left|\Tr(\rho_0 H)-\Tr(\rho_{\textup{Auf}} H)\right|,
\end{equation}
and the infidelity $1-\mathcal{F}^2(\rho_0,\rho_{\textup{Auf}})$ for different values of the magnetic field $h$. Here, $\rho_{\textup{Auf}}$ is the Aufbau state given in Eq.\ \eqref{Aufbaustate}, corresponding to the Hamiltonian in Eq.\ \eqref{eq:Heisenbergspinchain} and the probability distribution $\{p_i\}$ of the initial state given by Eqs. \eqref{eq:rank3overlaps} and \eqref{eq:rank5overlaps}.
The Uhlmann fidelity between the density matrices $\rho$ and $\sigma$ is defined as
\begin{equation}
\mathcal{F}(\rho,\sigma)
=\Tr\sqrt{\sqrt{\rho}  \sigma  \sqrt{\rho}}.
\end{equation} 
Note that both the energy difference and the infidelity are initially substantial, indicating that the randomly generated initial mixed states are generally far from the ordered equilibrium state constructed by the theorem. 

The QLLG evolution is solved numerically using a truncated perturbative expansion in $\kappa$; see the Supplementary Materials \cite{Supplementary Materials}. Alternative numerical methods for solving the QLLG have also been developed in \cite{azimi2025numerical, mirzaei2025lrei}. The upper panels in Fig.~\ref{fig:population and final error} shows the probability populations
\begin{equation}
P_i(t)=\bra{E_i}\rho(t)\ket{E_i},
\end{equation}
of the energy eigenstates $\ket{E_i}$ during the QLLG evolution. The dynamics exhibits a systematic redistribution of population toward lower-energy levels, leading to an increasingly ordered diagonal state in which lower-energy eigenstates acquire larger occupation probabilities. As seen in panels (a) and (b) in Fig.~\ref{fig:population and final error}, the initial populations are distributed across all energy levels, with relatively small occupation probabilities of each level, $i$. As the evolution proceeds, the populations progressively evolve toward lower energy eigenvalues and becomes finite only for the lowest $l=3, 5$ energy levels, where $l$ is the rank of the chosen initial states. 
This progressive emergence of population on the lowest-energy levels is precisely the energy-ordering mechanism predicted by the theorem.

To quantitatively verify the asymptotic ordering established by the theorem of this paper, the lower panels in Fig.~\ref{fig:population and final error} shows the final residual energy error
\begin{equation}
\label{eq:energyerror}
\varepsilon(\rho_{\mathrm{final}},\rho_{\textup{Auf}} ) =\left|
\Tr(\rho_{\mathrm{final}}H)
-
\Tr(\rho_{\mathrm{Auf}}H)
\right|,
\end{equation}
together with the infidelity
\begin{equation}
\label{eq:infidelityerror}
1-\mathcal{F}^2(\rho_{\mathrm{final}},\rho_{\mathrm{Auf}})
\end{equation}
for different values of the magnetic field $h$. Here, a small infidelity indicates that the final density matrix is close to the Aufbau state, i.e., the two density matrices have a high degree of similarity. Both quantities in Eqs. \eqref{eq:energyerror} and \eqref{eq:infidelityerror} remain small throughout the parameter range, demonstrating agreement between the numerically obtained asymptotic state and the ordered stationary state built by the theorem. As shown in the Supplementary Materials \cite{Supplementary Materials}, the convergence is slower for higher-rank initial states. Therefore, a longer time scale for the rank-$5$ case is used in panels (b) and (d), compared to the time scales for the rank-$3$ case in panels (a) and (c). The smaller infidelity and residual energy error observed in panel (d) reflect the longer evolution time, which allows the population distribution to converge more closely to the Aufbau state. Overall, the results provide direct numerical confirmation of the theorem and of the emergent energy-ordered population structure induced by the QLLG dynamics, and they are fully compatible with the analysis of the pure state case in \cite{Ataei2026}.

As a final comment, we note that our numerical analysis reveals a direct connection between the convergence of the QLLG dynamics and the spectral structure of the Hamiltonian. The relevant energy gaps control both the relaxation rate and the asymptotic accuracy: large gaps lead to faster relaxation, whereas small gaps generate slower convergence. Accordingly, the residual energy error and infidelity decrease as the relevant spectral gaps increase, consistent with the convergence behavior observed in the pure-state setting \cite{Ataei2026}. A further subtlety arises in the presence of energy degeneracies. When degenerate eigenspaces are present, the Aufbau-ordered state $\rho_{\mathrm{Auf}}$ is not uniquely specified at the level of its eigenvectors, since arbitrary unitary rotations within a degenerate eigenspace leave the energy unchanged. Consequently, two states can have nearly identical energies while exhibiting a finite infidelity. This explains the behavior observed in the lower panels of Fig.~\ref{fig:population and final error}. Although the residual energy error remains small across the parameter range, the infidelity exhibits higher values at some points associated with substantially low energy gaps. Further details are provided in the Supplemental Materials \cite{Supplementary Materials}. 

\textit{Conclusion}--We have investigated the dynamics of arbitrary mixed quantum states under the time evolution of the QLLG equation. By means of a stability analysis, we prove that QLLG drives generic mixed states toward a unique asymptotic configuration that is diagonal in the energy eigenbasis. The resulting state is obtained by arranging the probability spectrum of the initial density matrix in decreasing order and assigning it to the energy levels in increasing order, such that higher probabilities correspond to lower energy levels, providing a quantum analogue of the Aufbau principle.
We consider Heisenberg spin chains as a model system for numerical simulations of our analytical finding. The simulations confirm the theory and reveal a clear connection between the spectral structure of the Hamiltonian and the efficiency of the ordering process. In particular, both the convergence rate and the asymptotic accuracy are governed by the relevant energy gaps.
These results establish QLLG as a mechanism for dissipative ordering of energy populations and low-energy subspace preparation in mixed-state quantum systems. Beyond its connection to magnetic dynamics, the theory provides a general framework for nonlinear quantum relaxation in which spectral properties directly control both convergence speed and asymptotic performance. This opens avenues for applications in quantum optimization, state engineering, and dissipative quantum information processing.

\textit{Acknowledgment}--O.E. acknowledges financial support from the Swedish Research Council (VR) and the Knut and Alice Wallenberg Foundation (KAW).
O.E. also acknowledges support from the Wallenberg Initiative Materials Science (WISE), funded by the Knut and Alice Wallenberg Foundation, for support, as well as support from STandUPP, the ERC (FASTCORR project) and eSSENCE. The computations are enabled by resources provided by the National Academic Infrastructure for Supercomputing in Sweden (NAISS), partially funded by the Swedish Research Council (VR).

\bibliographystyle{apsrev4-2}

\section{Supplementary Materials}

\setcounter{section}{0}

\subsection*{A. Energy minimization on the isospectral manifold} 
The QLLG dynamics preserves the spectrum of the density matrix. Therefore, the evolution is restricted to the isospectral manifold
\begin{equation}
\mathcal{M}_{\rho_0}
=
\left\{
U\rho_0U^\dagger
\,:\,
U\in U(d)
\right\},
\end{equation}
where $d$ is the dimension of the Hilbert space and $\rho_0$ is the rank-$l$ initial density matrix of the QLLG dynamics. We prove that the asymptotic Aufbau state is precisely the unique minimizer of the energy functional
\begin{equation}
E[\rho]
=
\operatorname{Tr}(\rho H)
\end{equation}
over $\mathcal{M}_{\rho_0}$.
\vspace{2mm}
To prove this, let us assume that $\rho$ is a local minimizer of $E[\rho]$ on $\mathcal{M}_{\rho_0}$. Define the small perturbation of $\rho$ in $\mathcal{M}_{\rho_0}$ as follows
\begin{equation}
\rho(\varepsilon)
=
e^{\varepsilon A}
\rho
e^{-\varepsilon A},
\end{equation}
where $A$ is an arbitrary anti-Hermitian matrix. Since $\rho$ is a local minimizer, one must have
\begin{equation}
\operatorname{Tr}
\!\left(
e^{\varepsilon A}
\rho
e^{-\varepsilon A}
H
\right)
\ge
\operatorname{Tr}(\rho H)
\end{equation}
for sufficiently small $\varepsilon$. Differentiating at $\varepsilon=0$ gives
\begin{equation}
0
=
\frac{d}{d\varepsilon}
\operatorname{Tr}
\!\left(
e^{\varepsilon A}
\rho
e^{-\varepsilon A}
H
\right)
\Big|_{\varepsilon=0}.
\end{equation}
Using
\begin{equation}
\frac{d}{d\varepsilon}
\left(
e^{\varepsilon A}
\rho
e^{-\varepsilon A}
\right)_{\varepsilon=0}
=
[A,\rho],
\end{equation}
we obtain
\begin{equation}
0
=
\operatorname{Tr}
\!\left(
[A,\rho]H
\right).
\end{equation}
Hence, by the cyclicity of the trace, we have
\begin{equation}
\operatorname{Tr}
\!\left(
[\rho,H]A
\right)=\operatorname{Tr}
\!\left(
[A,\rho]H
\right)
=0.
\label{eq:criticalpoint}
\end{equation}
Multiplying  Eq.\eqref{eq:criticalpoint} by $i$, implies that the equation holds for  any Hermitian matrix $i A$. Hence, the linearity of the trace implies that
\begin{equation}
\operatorname{Tr}
\!\left(
[\rho,H]A
\right)
=
0
\end{equation}
holds for all matrices $A$, which is possible if and only if
\begin{equation}
[\rho,H]=0.
\label{eq:commutingcondition}
\end{equation}
Hence, every local minimizer of the constrained optimization problem must commute with the Hamiltonian.
\vspace{2mm}
Equation~(\ref{eq:commutingcondition}) implies that $\rho$ and $H$ can be simultaneously diagonalized. Using the energy eigenvalue decomposition 
\begin{equation}
H
=
\sum_{i=1}^{d}
E_i
|E_i\rangle\langle E_i|,
\qquad
E_1\le E_2\le\cdots\le E_d,
\end{equation}
and denoting the conserved eigenvalues of the initial density matrix $\rho_0$ by
\begin{equation}
p_1\ge p_2\ge \cdots \ge p_d,
\end{equation}
where only the first $l$ probabilities are nonzero, 
every stationary point has the form
\begin{equation}
\rho_\pi
=
\sum_{i=1}^{d}
p_{\pi(i)}
|E_i\rangle\langle E_i|,
\end{equation}
where $\pi$ is a permutation with corresponding energy
\begin{equation}
E[\rho_\pi]
=
\sum_{i=1}^{d}
p_{\pi(i)}E_i.
\label{eq:permenergy}
\end{equation}
Consider two permutations $\pi$, such that
\begin{equation}
E_i < E_j, \quad p_{\pi(i)} < p_{\pi(j)},
\end{equation}
and $\pi'$ coincides with $\pi$ for all indices except in two places $i,j$ as follows
\begin{align}
   \pi'(i) = \pi(j), \quad \pi'(j) = \pi(i).
\end{align}
  Then, the average energy difference satisfies 
\begin{equation}
\begin{aligned}
 \Tr(\rho_{\pi'} H) - \Tr(\rho_{\pi} H)
&=
(p_{\pi'(i)}E_i+p_{\pi'(j)}E_j)
-
(p_{\pi(i)}E_i+p_{\pi(j)}E_j)
\\
&=
(p_{\pi(j)}-p_{\pi(i)})(E_i-E_j)
<
0.
\end{aligned}
\end{equation}
Thus, any configuration in which a larger occupation is assigned to a higher energy is not a minimum, since exchanging the two occupations lowers the energy. In other words, the unique minimum is obtained when the lowest-energy eigenstate occupies the largest probability occupation, the second-lowest energy eigenstate occupies the second-largest eigenvalue, and so forth
\begin{equation}
\rho_{\rm Auf}
=
\sum_{i=1}^{d}
p_i
|E_i\rangle\langle E_i|,
\end{equation}
proving the Aufbau principle 
\begin{equation}
\rho_{\rm Auf}
=
\arg\min_{\rho\in\mathcal \mathcal{M}_{\rho_0}}
\operatorname{Tr}(\rho H)
.
\end{equation}


\subsection*{B. Numerical integration via a $\kappa$-series expansion}

In this section, we explain briefly how we carried out the simulation to solve the QLLG equation
\begin{equation}
\dot{\rho}
=
\frac{i}{\hbar}[\rho,H]
+
i\kappa[\rho,\dot{\rho},]
\label{eq:qllg_num}
\end{equation}
in the following steps.

\vspace{2mm}

\noindent
\textbf{Short-time parametrization.}
Consider a timestep of size $\Delta t$ and introduce the dimensionless variable
\begin{equation}
\tau=\frac{t-t_n}{\Delta t},
\qquad
0\le\tau\le1,
\end{equation}
where $t_n$ denotes $n$th time step. Then, we can expand the density matrix as the power series in $\kappa$,
\begin{equation}
\rho(t_n +\tau \Delta t)
=
\sum_{j=0}^{\infty}
\kappa^j
\rho^{(j)}.
\label{eq:kappaexpansion}
\end{equation}
Substituting Eq.~\eqref{eq:kappaexpansion} into Eq.~\eqref{eq:qllg_num} and collecting equal powers of $\kappa$ yields a hierarchy of equations, where the zeroth-order contribution reads
\begin{equation}
\partial_t \rho^{(0)}
=
\frac{i}{\hbar}
[\rho^{(0)},H],
\label{eq:zerothorder}
\end{equation}
that corresponds to ordinary unitary evolution. The higher-order equations obey
\begin{equation}
\partial_t \rho^{(j+1)}
=
i
\sum_{a+b=j}
\Big[
\rho^{(a)},
\partial_t \rho^{(b)}
+
\frac{i}{\hbar}
[\rho^{(b)},H]
\Big],
\label{eq:recursive}
\end{equation}
for $j \geq 0$. Equation~\eqref{eq:recursive} depends only on lower-order terms and therefore can be evaluated recursively.

\vspace{2mm}

\noindent
\textbf{Polynomial representation.} Taking the integral of Eq.~\eqref{eq:recursive}, we derive
\begin{equation}
\rho^{(j+1)}(\tau)
=
i\Delta t
\int_0^\tau
d\tau'
\sum_{a+b=j}
\Big[
\rho^{(a)}(\tau'),
\partial_{\tau'}\rho^{(b)}(\tau')
+
\frac{i\Delta t}{\hbar}
[\rho^{(b)}(\tau'),H]
\Big].
\label{eq:integralrecursion}
\end{equation}
Since every term is represented as a polynomial in $\tau$, the integration can be carried out such that each coefficient $\rho^{(j)}(\tau)$ is represented as a polynomial in $\tau$,

\begin{equation}
\rho^{(j)}(\tau)
=
\sum_{m=0}^{M_j}
A_m^{(j)}
\tau^m ,
\end{equation}

with matrix coefficients $A_m^{(j)}$. 

\vspace{2mm}

\noindent
\textbf{Single-step propagator.} Using the above procedure QLLG state can be approximated as
\begin{equation}
\rho(t_n+\tau \Delta t)
=
\sum_{j=0}^{n_{\mathrm{max}}}
\kappa^j
\rho^{(j)}(\tau).
\label{eq:singleprop}
\end{equation}
at time step $t_{n+1}$ up to order $n_{\mathrm{max}}$, where we consider
$n_{\mathrm{max}}=4$ for our simulation.

\vspace{2mm}

\subsection*{C. Convergence time from coherence decay}

The asymptotic Aufbau state established in the main text is diagonal in the energy eigenbasis. Consequently, convergence toward the asymptotic state is governed by the suppression of off-diagonal coherences. Let
\begin{equation}
H|E_i\rangle = E_i |E_i\rangle.
\end{equation}
To determine the convergence rate, we linearize Eq.~(\ref{eq:qllg_num}) around the asymptotic fixed point $\rho(\infty)$, i.e., \begin{equation}
\label{eq:perturbdensity}
\rho(t)
=
\rho(\infty)
+
\delta\rho(t),
\end{equation}
where $\delta\rho$ is a small perturbation. Since $\rho(\infty)$ is stationary,
\begin{equation}
[\rho(\infty),H]=0,
\end{equation}
and therefore it is diagonal in the energy basis,
\begin{equation}
\rho(\infty)
=
\sum_i
p_i
|E_i\rangle\langle E_i|.
\end{equation}
Substituting Eq. \eqref{eq:perturbdensity} 
into Eq.~(\ref{eq:qllg_num}) gives \begin{equation}
\dot{\delta\rho}
=
\frac{i}{\hbar}
[\rho(\infty)+\delta\rho,H]
+
i\kappa
[\rho(\infty)+\delta\rho,\dot{\delta\rho}].
\end{equation}
Keeping only terms linear in $\delta\rho$ yields
\begin{equation}
\dot{\delta\rho}
=
\frac{i}{\hbar}
[\delta\rho,H]
+
i\kappa
[\rho(\infty),\dot{\delta\rho}],
\label{eq:linearizedQLLG}
\end{equation}
where all quadratic terms of order
$\mathcal{O}(\delta\rho,\dot{\delta\rho})$
have been neglected.

Taking matrix elements of Eq.~(\ref{eq:linearizedQLLG}) in the energy basis gives
\begin{equation}
\dot{\delta\rho}_{ij}
=
\frac{i}{\hbar}
(E_j-E_i)
\delta\rho_{ij}
+
i\kappa
(p_i-p_j)
\dot{\delta\rho}_{ij},
\end{equation}
where $\rho_{ij} = \bra{E_i} \rho(t) \ket{E_j}.$ Therefore,
\begin{equation}
\label{eq:modeequation}
\dot{\delta\rho}_{ij}
=
\left(
-\Gamma_{ij}
-
i\Omega_{ij}
\right)
\delta\rho_{ij},
\end{equation}
where
\begin{equation}
\Gamma_{ij}
=
\frac{
\kappa
(p_i-p_j)
(E_i-E_j)
}{
\hbar
\left[
1+\kappa^2(p_i-p_j)^2
\right]
},
\label{eq:gammaij}
\end{equation}
and
\begin{equation}
\Omega_{ij}
=
\frac{
(E_i-E_j)
}{
\hbar
\left[
1+\kappa^2(p_i-p_j)^2
\right]
}.
\label{eq:omegaij}
\end{equation}
This gives rise to the solution
\begin{equation}
\delta\rho_{ij}(t)
=
\delta\rho_{ij}(0)
e^{-\Gamma_{ij}t}
e^{-i\Omega_{ij}t}.
\label{eq:modeSolution}
\end{equation}
Since $\rho(\infty)$ is diagonal, by Eq. \eqref{eq:perturbdensity} we obtain that the off-diagonal matrix elements of $\rho$ coincide with those of $\delta\rho$, and thus
\begin{equation}
\rho_{ij}(t)
=
\rho_{ij}(0)
e^{-\Gamma_{ij}t}
e^{-i\Omega_{ij}t},
\qquad i\neq j.
\label{eq:coherenceEnvelope}
\end{equation} For a given threshold $\varepsilon$, the decay time associated with a single coherence mode is obtained from
\begin{equation}
|\rho_{ij}(t)|=\varepsilon.
\end{equation}
that yields 
\begin{equation}
\tau_{ij}
=
\frac{
\log
\left(
\frac{|\rho_{ij}(0)|}{\varepsilon}
\right)
}{
\Gamma_{ij}
},
\label{eq:tauij}
\end{equation}
following Eq. \eqref{eq:coherenceEnvelope}. The overall convergence time is then controlled by the slowest-decaying coherence,
\begin{equation}
\tau_{\mathrm{conv}}
=
\max_{i\neq j}
\tau_{ij},
\label{eq:tauconv}
\end{equation}
which gives
\begin{equation}
\tau_{\mathrm{conv}}
= \max_{i\neq j}
\left[
\frac{
\hbar
\left(
1+\kappa^2(p_i-p_j)^2
\right)
}{
\kappa
|p_i-p_j|
|E_i-E_j|
}
\log
\left(
\frac{|\rho_{ij}(0)|}{\varepsilon}
\right)
\right],
\label{eq:fulltau}
\end{equation}
using Eq. \eqref{eq:gammaij}. In the weak-damping regime $
\kappa\ll1,$ one can obtain
\begin{equation}
\Gamma_{ij}
\simeq
\frac{\kappa}{\hbar}
(p_i-p_j)
(E_i-E_j),
\end{equation}
and therefore
\begin{equation}
\tau_{\mathrm{conv}}
\simeq
\max_{i\neq j}
\frac{
\hbar
\log\left(
\dfrac{|\rho_{ij}(0)|}{\varepsilon}
\right)
}{
\kappa
|p_i-p_j|
|E_i-E_j|
}.
\label{eq:weaklimit}
\end{equation}
Eq. ~(\ref{eq:weaklimit}) indicates that convergence is controlled by energy gaps as well as density-matrix eigenvalue gaps. Small spectral gaps therefore produce slow relaxation, which is also observed in numerical simulations near level crossings and quasi-degenerate spectra.
 
\end{document}